\documentclass[12pt,preprint]{aastex}

\usepackage{graphics,epsfig}
\usepackage{natbib}

\def\cM{{\cal M}}
\def\cO{{\cal O}}
\def\cC{{\cal C}}
\def\rO{{{\rm O}_2}}

\def\spose#1{\hbox to 0pt{#1\hss}}
\def\lta{\mathrel{\spose{\lower 3pt\hbox{$\mathchar"218$}}
     \raise 2.0pt\hbox{$\mathchar"13C$}}}
\def\gta{\mathrel{\spose{\lower 3pt\hbox{$\mathchar"218$}}
     \raise 2.0pt\hbox{$\mathchar"13E$}}}

\def\figure#1#2 {\par{\narrower\noindent {\bf Fig. #1}
   \hskip 2mm #2\par}\bigskip\noindent}
\def\table#1#2 {\par{\narrower\noindent {\bf Tab. #1}
   \hskip 2mm #2\par}\bigskip\noindent}

\slugcomment{Version: \enskip \today}

\shorttitle{Fractal and Multifractal Analysis of Earth's Rise of Oxygen}

\shortauthors{Kumar, Cuntz \& Musielak}

\begin{document}


\title{
Fractal and Multifractal Analysis of the Rise of Oxygen \\
in Earth's Early Atmosphere
}

\author{Satish Kumar}
\author{Manfred Cuntz}
\author{Zdzislaw E. Musielak}

\affil{Department of Physics, University of Texas at Arlington}
\affil{Arlington, TX 76019;}
\email{satish.kumar@mavs.uta.edu; cuntz@uta.edu; zmusielak@uta.edu}

\begin{abstract}
The rise of oxygen in Earth's atmosphere that occurred 2.4 to 2.2 billion 
years ago is known as the Earth's Great Oxidation, and its impact on the
development of life on Earth has been profound.  Thereafter, the increase
in Earth's oxygen level persisted, though at a more gradual pace.  The
proposed underlying mathematical models for these processes are based
on physical parameters whose values are currently not well-established
owing to uncertainties in geological and biological data.  In this paper,
a previously developed model of Earth's atmosphere is modified by adding
different strengths of noise to account for the parameters' uncertainties.
The effects of the noise on the time variations of oxygen, carbon and methane
for the early Earth are investigated by using fractal and multifractal analysis.
We show that the time variations following the Great Oxidation cannot
properly be described by a single fractal dimension because they exhibit
multifractal characteristics.  The obtained results demonstrate that the
time series as obtained exhibit multifractality caused by long-range
time correlations.
\end{abstract}

\keywords{Earth's Great Oxidation, Hurst exponent, fractal dimension, 
multifractal analysis.}


\section{INTRODUCTION}

Fractal methods are well-suited for describing self-similarity, jaggedness 
and correlation of data sets.  Typically, the data is characterized by the 
so-called fractal dimension, which can be evaluated by 
finding how the data fills its embedding space [1].  The data can be 
plotted and then the jaggedness of such graph can be compared to a straight line, 
leading to the box counting method or correlation integral methods.  Another way 
is to treat the data as discrete and to compare it to uncorrelated noise,
like in the well-known Hurst exponent method [2].  The fractal methods are 
generalized by multifractal methods that have different scaling moments for
different moments or for different magnitudes of fluctuations.  However, if
a scaling component depends on the scale, there is a crossover, and the
multifractal analysis must be separately applied to ranges of small and large
scales.  Typically, a singularity spectrum [3,4] is computed to determine 
multifractal characteristics of time series data.

The fractal and multifractal methods have been extensively applied to different 
problems in natural sciences, engineering, medicine, and the social sciences
[3,4,5,6,7,8,9,10].  However, these methods have only been sparsely applied
within the recently established field of astrobiology or its terrestrial counterpart,
i.e., biogeology / geobiology, though the study by [11]
on exoplanetary life detectability represents a notable exception.  In this paper,
we pursue another applications, a topic of great importance, namely, the rise of
oxygen in Earth's early atmosphere.  It consists of the Earth's Great Oxidation,
followed by a more gradual increase of Earth's atmospheric oxygen level.

In the following, we describe the problem in detail, and also explain
why we chose the fractal and multifractal methods for this study.   
The Earth's Great Oxidation occurred 2.4 to 2.2 billion years ago and it 
had a significant impact on Earth's atmospheric physics and chemistry and, 
furthermore, entailed profound implications for the evolution of life.  
During this time span, the atmospheric oxygen concentration rose from 
less than 10$^{-5}$ of the present atmospheric level (PAL) to more than
0.01 PAL and possibly above 0.1 PAL [12,13,14,15].  Thereafter, a more gradual
increase of the oxygen level occurred, eventually leading to today's level.
The rise of oxygen is still a topic of lively discussion owing to its 
wide ramifications regarding terrestrial and planetary astrobiology;
specifically, it is still unclear whether Earth's Great Oxidation occurred
relatively smoothly or exhibited big jumps, i.e., akin to a yoyo
model\footnote{Yoyo atmosphere is a term introduced by Ohmoto et al. (2006) [16]
to refer to hypothetical oxygen in Earth's atmosphere prior to the Great
Oxidation.  However, in the following this term is used as done by 
Cuntz et al. (2009) [17] to describe possible significant ups and downs of
early Earth's atmospheric oxygen amounts, which could also have occurred during
or after the Great Oxidation.} (see [18]).  Moreover, the entire
sequence of oxygen increases, including the increase during the Proterozoic eon
after the Great Oxidation, could have occurred relatively smoothly or by
exhibiting intricate mathematical structures, which if present can only
be uncovered through multifractal analysis.

Information on the rise of oxygen, especially the Great Oxidation, is
in part based on studies of atmospheric sulphur chemistry, including
analyses of multiple sulphur isotopes, as, e.g., the isotopic ratios
between $\Delta^{32}$S, $\Delta^{34}$S, and $\Delta^{36}$S; see work by
Farquhar et al. (2000) [19], Ohmoto et al. (2006) [16], Farquhar et al.
(2007) [20], and Johnston (2011) [21] (in analogy using isotopic ratios
to estimate the age of the Earth [22]).
These studies indicate the patterns of
increase of early Earth's oxygen levels during different time intervals,
including possible fluctuations.  Studies providing additional insight into
the long-standing debate regarding O$_2$ in Earth's early atmosphere include
the work by Claire et al. (2006) [23], Lyons (2007) [24], Kaufman et al. (2007)
[25], Kump (2008) [26], Balk et al. (2009) [27], Frei et al. (2009) [28],
Freund et al. (2010) [29], Freund (2011) [30], and Flannery \& Walter (2012) [31].

A set of nonlinear equations describing the time evolution of oxygen,
methane and carbon in the early Earth was originally proposed and solved
by Goldblatt et al. (2006) [32], thereafter GLW06, who encountered
bistability in the system equations, which in their model represent
the ancient Earth's low-oxygen state ($\lta 10^{-5}$~PAL) and the
high-oxygen state ($5 \times 10^{-3}$~PAL).  Subsequent work by
Cuntz et al. (2009) [17], thereafter CRM09, on the nonlinear set of
equations further explored this system by replacing the original step
function (GLW09) representing the reductant input rate by more realistic
functions (i.e., exponential decay function, logistic decay function)
with and without Gaussian white noise.  Based on the transition stability
analysis for the system equations, CRM09 considered a set of non-autonomous,
nonlinear differential equations and furthermore inspected the Lyapunov
exponents [33,34,35].  CRM09 found that the equations do not show
chaotic behavior and that the rise of oxygen during the Great Oxidation
occurred relatively smoothly.

Previous models of Earth's Great Oxidation and the subsequent oxygen
increase during the early Proterozoic eon have been based on a set
of physical parameters that were determined
using the available geological data as summarized by GLW06 and CRM09.
However, it is well-known that this set of parameters is not unique
because of potentially large uncertainties in the data.  Therefore, 
it is the aim of the present work to expand those previous studies 
by adopting other sets of parameters and also analyzing the obtained 
results by using fractal and multifractal techniques.  Specifically, 
we will use the standard Hurst exponent [2] and the fractal 
dimension related to it [1], as well as the Multifractal 
Detrended Fluctuation Analysis (MFDFA) [34], which
was originally developed for non-stationary time series and required 
the so-called generalized Hurst exponents for computing a width of 
singularity spectrum (WSS).  Our choice of using the MFDFA is motivated 
by the fact that our numerical solutions constitute non-stationary
time series.  

Our studies are performed for different strengths of white Gaussian
noise, which is added to the system to account for possible variations 
in the physical parameters as well as for uncertainties in the adopted 
values of the parameters.  We consider different levels of noise in the 
system and investigate their effects on the rise of oxygen as well as 
on the associated time variations of atmospheric methane and carbon in
the Earth's surficial environment.  The effects of the noise are studied
by  performing fractal and multifractal analyses of our numerical results.  
We will be able to demonstrate that no single fractal dimension 
can be used to describe time variations of oxygen, carbon and methane 
because they exhibit multifractal characteristics due to long-range 
correlations of the small and large fluctuations in the time series.  

Our paper is structured as follows:  In \S 2, we describe the system 
equations considered in our study, the numerical method of solution, 
and the methods for performing the fractal and multifractal analysis.  
In \S 3, we present our results and discussion.  Our conclusions are 
given in \S 4.


\section{FORMULATION AND ANALYSIS TECHNIQUES}

\subsection{Original Governing Equations}

The set of equations was originally given by GLW06 and was subsequently
revisited by CRM09.  It encompasses a simplified model of Earth's global 
redox system, representing the atmosphere, ocean, and crust.  Concerning 
the atmosphere and ocean, the number of moles of methane $\cM$ (CH$_4$) 
and oxygen $\cO$ (O$_2$) are calculated.  Furthermore, with respect to 
the ocean floor, the amount of buried organic carbon $\cC$ in the crust 
is also computed.  This leads to the following system of equations:
\begin{equation}
\frac{d\cM}{dt} \ = \ \frac{1}{2}\Omega_{(\rO)}(N+r) - \frac{1}{2}
\Psi_{(\rO)}\cM^{0.7} - s\cM - \frac{1}{2}\Omega_{(\rO)}
\Bigl[\beta(N+r)-w\cC\Bigr]\ ,
\label{eq1}
\end{equation}

\begin{equation}
\frac{d\cO}{dt} \ = \ \Omega_{(\rO)}N - (1-\Omega_{(\rO)})r - 
\Psi_{(\rO)}\cM^{0.7} - s\cM - (1-\Omega_{(\rO)})
\Bigl[\beta(N+r)-w\cC\Bigr]\ ,
\label{eq2}
\end{equation}
and
\begin{equation}
\frac{d\cC}{dt} \ = \ \beta (N+r) - w\cC\ .
\label{eq3}
\end{equation}

This set of differential equations is coupled and nonlinear because of the
term $\cM^{0.7}$ as well as the functions $\Omega_{(\rO)}$ and $\Psi_{(\rO)}$.
Note that $\Omega_{(\rO)}$ is given as $\Omega_{(\rO)} = (1-\gamma)(1-\delta)$
with $\gamma=\cO/(d_\gamma+\cO)$ and $\delta=\cO/(d_\delta+\cO)$, whereas 
$\Psi_{(\rO)}$ is given as $\Psi_{(\rO)} = 10^\epsilon$ with $\epsilon = 
a_1\varphi^4+a_2\varphi^3+a_3\varphi^2+a_4\varphi+a_5$, $\varphi=\log\cO$, 
$a_1=0.0030$, $a_2=-0.1655$, $a_3=3.2305$, $a_4=-25.8343$ and $a_5=71.5398$.

The different terms, parameters, and variables of Eqs. (1) to (3) require
further detailed explanations.  The first term on the RHS of ${d\cM}/{dt}$ 
and the corresponding terms of ${d\cO}/{dt}$ denote the net primary productivity 
for oxygenic photosynthesis $N$ and the net input of reductant to the surface 
$r$, an electron donor for anoxygenic photosynthesis, which both depend on 
the time-variable oxygen concentration $\cO$.  Both GLW06 and CRM09 took
$N = 3.75 \times 10^{15}$ mol O$_2$~yr$^{-1}$.  However, for $r = r(t)$,
for which GLW06 chose a step function, exhibiting a downward jump from
$3 \times 10^{11}$ to $7.5 \times 10^{10}$ mol O$_2$ equiv. yr$^{-1}$ at
2.4 Gyr, CRM09 adopted different functions, including functions incorporating
statistical noise.  This allowed them to consider the impact of different
plausible reductant functions on early Earth's atmospheric chemistry.

The work by GLW06 and CRM09 considered the dominant processes of the marine
biosphere related to oxygenic photosynthesis and represented by an assumed
net primary productivity of $N$ mol O$_2$~yr$^{-1}$, implying that the
total organic carbon produced is given as $(N+r)$.  The fraction of this
amount consumed by heterotrophic respirers is $\gamma=\cO/(d_\gamma+\cO)$
with $d_\gamma = 1.36 \times 10^{19}$~mol.  The fraction of the methane
produced that is consumed by methanotrops is given as $\delta=\cO/(d_\delta 
+ \cO)$, where $d_\delta = 2.73 \times 10^{17}$~mol.  Therefore, the fraction
of the oxygen and methane  produced that reaches the atmosphere is given by
$\Omega_{(\rO)} \ = \ (1-\gamma)(1-\delta)$.  The next two terms on the RHS
of Eqs. (1) and (2) correspond to the atmospheric methane oxidation reaction.
Its rate can be obtained by fitting  the results of detailed photochemical
models, including the introduction of  $\Psi_{(\rO)}$, which constitutes
a polynomial function with respect to the  amount of oxygen.

Once there is sufficient oxygen, Earth's ozone layer forms, thus providing
an effective shield against UV radiation.  There are other terms in
Eqs. (1) and (2) related to sinks and sources of atmospheric methane $\cM$.
Methane can be nullified by oxidation, but it can also be lost through hydrogen
escape since methane is considered to be the sole source of hydrogen in
the upper atmosphere with $s = 3.7 \times 10^{-5}$ yr$^{-1}$ (GLW06) as a
limiting flux constant.  Detailed studies on the loss of hydrogen have
been given by Hunten (1982) [37], Catling et al. (2001) [13], Zahnle et al. (2013)
[38], among others.  The final terms on the RHS of Eqs. (1) to (3) are due
to the amount of buried organic carbon, which depends on the fraction
$\beta$, taken as $2 \times 10^{-3}$, of the total organic carbon $(N+r)$
that is produced.  The rate of organic carbon weathering is $w\cC$, where
the bulk weathering rate $w$, given as $w = 6 \times  10^{-9}$ yr$^{-1}$,
is set by analogy to present conditions and $\cC$ is the crustal organic
carbon; see GLW06 and CRM09 for further details.

GLW06 obtained steady state solutions for methane and carbon given 
as $\cM = r/s$ and $\cC = \beta (N+r)/w$, respectively.  In case of oxygen,
the steady state solution required obtaining the roots of a more complicated
equation.  In GLW06's model, the solutions for oxygen, methane and carbon,
exhibit bistability.  This result, as previously pointed out by Kasting (2006) [18],
arguably offers crucial insights into the bistable nature of Earth's atmosphere,
especially pertaining to the Great Oxidation, considering that the latter
constitutes a switch to the high-oxygen (more than $5 \times 10^{-3}$ PAL)
steady state from a previous low-oxygen (less than $10^{-5}$ PAL)
steady state that existed for at least 300 million years even after the
onset of oxygenic photosynthesis.

\subsection{Modified Governing Equations with Noise}

The main objective of our work is to add white Gaussian noise to the governing 
equations, i.e., Eqs. (\ref{eq1}) to (\ref{eq3}), in order to investigate its 
effect on the rise of oxygen in the early Earth's atmosphere.  We therefore 
introduce the following three new variables:  $\eta_{\cM} = \sigma \kappa \cM$, 
$\eta_{\cO} = \sigma \kappa \cO$, and $\eta_{\cC} = \sigma \kappa \cC$, where 
$\sigma$ varies from $0.005$ to $0.010$, and $\kappa$ is a random number given 
by a normal distribution with a mean of 0 and a variance of 1.  Furthermore, 
$\cM$, $\cC$, and $\cO$ represent the amounts of methane, carbon, and oxygen, 
respectively, as before.  With the new variables added, Eqs. (\ref{eq1}) to 
(\ref{eq3}) read as
\begin{equation}
\frac{d\cM}{dt} \ = \ \frac{1}{2}\Omega_{(\rO)}(N+r) - \frac{1}{2}
\Psi_{(\rO)}\cM^{0.7} - s\cM - \frac{1}{2}\Omega_{(\rO)}
\Bigl[\beta(N+r)-w\cC\Bigr] + \eta_{\cM}\ ,
\label{eq4}
\end{equation}

\begin{equation}
\frac{d\cO}{dt} \ = \ \Omega_{(\rO)}N - (1-\Omega_{(\rO)})r - 
\Psi_{(\rO)}\cM^{0.7} - s\cM - (1-\Omega_{(\rO)})
\Bigl[\beta(N+r)-w\cC\Bigr] + \eta_{\cO}\ ,
\label{eq5}
\end{equation}
and
\begin{equation}
\frac{d\cC}{dt} \ = \ \beta (N+r) - w\cC + \eta_{\cC}\ .
\label{eq6}
\end{equation}

These equations are solved using the geological and Earth's atmospheric data
as discussed by GLW06 and CRM09 (see \S 2.1).  However, the new variables
$\eta_{\cM}$, $\eta_{\cO}$, and $\eta_{\cC}$ allow us to consider appropriate
levels of noise for the terms $d{\cM}/dt$, $d{\cO}/dt$, and $d{\cC}/dt$,
respectively, in the view of inherent uncertainties in the underlying
geological processes.

Equations (\ref{eq4}) to (\ref{eq6}) are solved for different values of 
the noise strength $\sigma$ by using the Euler-Maruyama method, which 
is suitable for the considered set of equations; additionally, the 
resulting numerical errors are small.  The obtained numerical results
are displayed in 3D phase space defined by the variables $\cO$, $\cC$
and $\cM$.  We refer to such displays of our results as phase portraits
whose main advantage is that they are able to show relationships between
the three variables, and, furthermore, allow to demonstrate how the
relationships change when the noise strength $\sigma$ is varied.
In order to determine how effectively the change occurs, we use the
standard Hurst exponent [2] and the fractal dimension related
to this exponent [1] as well as the so-called multifractal
technique, which in \S 2.3.2 will be described in detail.

\subsection{Analysis Techniques}

\subsubsection{Hurst Exponent and Fractal Dimension}

The standard Hurst exponent $h$ was originally introduced by Hurst (1951) [2], 
who demonstrated that data described by $h$ is anti-correlated if $0 \le h 
< 0.5$, uncorrelated if $h = 0.5$, and correlated if  $0.5 < h \le 1.0$.  
According to Mandelbrot (1982) [1], $h$ can be used to introduce the fractal 
dimension $D_h = \epsilon - h$, where $\epsilon$ is the embedding 
space, and $D_h$ is given as

\begin{mathletters}
\begin{eqnarray}
D_h = {1 \over {1-h}}    &  0   \leq h < 0.5   &  (\rm {anti-persistent}) \\
\label{eq7a}
D_h = 2.0                &  h   = 0.5          &  (\rm {non-persistent}) \\
\label{eq7b}
D_h = {1 \over h}        &  0.5 < h \leq 1.0   &  (\rm {persistent}) \ .
\label{eq7c}
\end{eqnarray}
\end{mathletters}
The above definition of $D_h$ is known to provide a smooth transition 
from anti-persistent to persistent including the point $h = 0.5$, 
and to be robust for measuring macro-trends in the data [4].

\subsubsection{Multifractal Technique}

We analyze the results of our numerical simulations by using the 
Multifractal Detrended Fluctuation Analysis (MFDFA), which is used to 
compute the so-called generalized Hurst exponents $h(q)$ and a width 
of singularity spectrum; note that the generalized Hurst exponents
$h(q)$ becomes the standard Hurst exponent $h$ if $q = 2$.  Let us 
now briefly describe the MFDFA method by following the notation and 
steps originally introduced by Kantelhardt et al. (2002) [36].  We begin
by calculating the profile for a given series $Y(i)$, where
$i = 1, 2,..., N$, using 
\begin{equation}
Y(i) = \sum_{k=1}^i [x_k - <x>]\ ,
\label{eq8}
\end{equation}

\noindent
profile into $N_s = N/s$ non-overlapping segments of equal length $s$.  
In general, $N$ may not be a multiple of $s$, thus in order to account 
for a small part remaining at the end of the series, we repeat the same 
procedure but this time by starting from the end of the series; note that 
reverting the series does not affect the multifractal behavior of the 
series as all its parameters remain the same.  Since the procedure is
used twice, $2 N_s$ segments are obtained.

We determine a possible local trend of the series by the least-square
polynomial fit of the series.  Thereafter, we calculate the variance by using
\begin{equation}
F^2 (\nu, s) = {1 \over s} \sum_{i=1}^s [Y((\nu-1)s + i) - y_{\nu}(i)]^2
\eqnum{9a}
\label{eq9a}
\end{equation}
for the segments $\nu$ = 1, 2,..., $N_s$, and
\begin{equation}
F^2 (\nu, s) = {1 \over s} \sum_{i=1}^s [Y(N - (\nu-N_s)s + i) - y_{\nu}(i)]^2 
\eqnum{9b}
\label{eq9b}
\end{equation}
\noindent
for the segments $\nu$ = $N_s + 1$,..., $2N_s$, with $y_{\nu} (i)$ being the 
fitting polynomial in segment $\nu$. 

With the variance known, we compute the moment dependent fluctuation function 
\begin{equation}
F_q (s) = \Big[ {1 \over {2N_s}} \sum_{\nu=1}^{2N_s} [F^2 (\nu, s)]^{q/2}
\Big]^{1/q}\ .
\eqnum{10a}
\label{eq10a}
\end{equation}
Note that for $q=2$, the above analysis becomes the detrended fluctuation analysis
(DFA).  However, for other values of $q$, the multifractal behavior is characterized
by different Hurst exponents at different scales, the so-called generalized Hurst
exponent, which requires the computation of the fluctuating function at different moments.   

The power law behavior is displayed in log--log plots of the fluctuating function 
$F_q (s) \sim s^{h(q)}$, where the generalized Hurst exponent $h(q)$ converges 
to the standard Hurst exponent $h$ of the series.  Since scales $s < 10$ and $s > 
N/4$ lead to systematic statistical errors, we take the values between them.  
Moreover, $F_q (s)$ diverges for $q=0$, which means that it cannot be calculated 
using Eq. (\ref{eq9a}, \ref{eq9b}); therefore, we compute it by using a logarithmic average 
given as
\begin{equation}
F_q (s) = \exp \Big[ {1 \over {4N_s}} \sum_{\nu=1}^{2N_s} \ln [F^2 (\nu, s)]
\Big] \sim s^{h(0)}\ .
\eqnum{10b}
\label{eq10b}
\end{equation}

\noindent
The generalized Hurst exponent is related to the classical scaling exponent $\tau (q)$ 
through the following relation $\tau (q) = q h(q) - 1$.

There is also another way of characterizing the multifractal series.  It involves
introducing  the so-called singularity spectrum $f(\xi)$, which is related to $\tau (q)$
via a Legendre transform, i.e., $\xi = \tau^{\prime} (q)$ and $f(\xi) = q \xi-\tau 
(q)$, with $\xi$ being the singularity strength or H\"older exponent; note that 
the WSS provides a measure of strength of  multifractal behavior, and that it is used
in this study to analyze the results of our numerical simulations.
%


\section{RESULTS AND DISCUSSION}

\subsection{Numerical Results}

We solve Eqs. (\ref{eq4}) through (\ref{eq6}) by using the Euler-Maruyama method. 
The input parameter $r(t)$, i.e., the net input of reductant to the surface, is
specified as given by the exponential function $r(t) = r_0 \exp (-\alpha t/t_0)$,
with $r_0 = 3.5 \times 10^{11}$ mol O$_2$ equiv. yr$^{-1}$, $\alpha = 1.4$, and
$t_0 = 1.0 \times 10^7$ yr.  In addition, we assume that the initial number 
of oxygen molecules is $10^8$ mol and use a step size of $10^{-3}$ yr.  Our numerical 
computations indicate time-dependent changes for the amounts of oxygen, methane and
carbon.  The changes of oxygen in time are presented in Fig.~1, which shows an
initially smooth, but steep rise within the first $10^3$ years, and then oscillatory
behavior, which is driven by the noise.  It is this oscillatory behavior, which
mostly occurs after the Great Oxidation, i.e., during the early Proterozoic eon,
we wish to focus on.  Our study will be pursued in the 3D phase space defined
by the variables describing carbon, oxygen and methane of the system.

The obtained phase portraits for different strengths of noise are shown in Figs.~2
and 3; note that our numerical results obtained for the first $10^3$ years have
been omitted.  The phase portraits show the relationships between the carbon, oxygen
and methane variables as well as the extent of the 3D phase space covered by these
relationships for the different levels of noise.  By inspecting the phase portraits,
we find that different strengths of noise correspond to remarkably different depictions
in the phase space.  However, there is not a clear correlation between the strength
of noise and the volume of the 3D phase space covered by the relationships.  Therefore,
as the next step we perform the fractal and multifractal analysis of our numerical
results.  

Through this analysis, we want to determine whether the resulting numerical time
series describing variations of carbon, oxygen and methane in our model of the early
Earth is a fractal system that requires a single exponent or fractal dimension,
or a multifractal system that is characterized by a continuous spectrum of exponents 
or the WSS.  The analysis will also allow us to identify the nature of multifractality,
i.e., whether it is due to probability distributions or long-range correlations.

\subsection{Generalized Hurst exponent and fractal dimension}

We compute the generalized Hurst exponent $h$ by using the MFDFA and taking $q = 2$, 
which means that the resulting $h$ is equivalent to the standard Hurst exponent
[2].  The values of $h$ obtained for different strengths of the noise 
introduced to the system are shown in the second column of Table~1.  It is seen that 
$h$ is not sensitive to the different values of noise strengths and that it is close 
to $0.5$, which implies that the numerically computed time series data is uncorrelated.  

Having obtained $h$, we use the results presented in \S 2.3 to calculate the fractal 
dimension $D_h$.  The computed values of $D_h$ are given in the third column of Table~1. 
It is seen that $D_h$ is largely independent of the strength of the noise, which
is an expected result because of the $D_h$ dependence on $h$.  This obviously implies
that our numerically computed time series of variations of carbon, oxygen and methane
in our model of the early Earth cannot be fully characterized by one
single exponent or fractal dimension.  Therefore, our main conclusions about 
the effects of noise on our numerical results must arise from the multifractal 
analysis and the concept of the WSS.

\subsection{Multifractal analysis}   

Following \S 2.3, we perform a multifractal analysis based on the MFDFA,
which was used to compute the WSS, where the WSS represents the distance
between two end points of the singularity spectrum, i.e., the first and
last values of the H\"older exponent $\xi$.  Note that the higher the
value of the WSS, the stronger is the multifractal characteristics of
the time series.  The computed values of the WSS are given in the
last column of Table~1.  

It is found that the value of the WSS is clearly the highest for the
strength of noise $\sigma = 0.005$ and the lowest for $\sigma = 0.010$.
There is an obvious trend indicating a negative slope between these
values in terms of a $\sigma$--WSS relationship, albeit the existence
of scatter, which is most evident regarding $\sigma = 0.008$, where
the value for the WSS is close to the minimum value at $\sigma = 0.010$.
In order to gauge the significance of the decline of the WSS values as
function of $\sigma$, we applied a two-tailed Spearman $r$ correlation
analysis (e.g., [39]).  The Spearman $r$ is identified
as $-0.71$ with a standard deviation of 0.11, hence confirming
the suspected functional decline.  Thus, the general trend of the
decreasing WSS with increasing values of $\sigma$ appears to be valid
for our numerically generated time series.

The observed trend is an indication that white Gaussian noise changes
the multifractal characteristics of our time series, which is contrary
to the fractal dimension that is about the same for all time series
as shown in  the third column of Table~1.  Therefore, the inclusion
of the noise  makes a significant contribution to the overall behavior
of oxygen and methane in our model of Earth's early atmosphere and to
carbon present in early Earth's surficial environment.  Additionally,
our results show that this contribution can only be determined by 
using the multifractal analysis, which is in agreement with the
earlier finding [40,41] that a true signal is distinguished from
superimposed noise if MFDFA is employed in natural time [42].

The above analysis was performed for the white Gaussian noise with 
$\sigma$ varying from $0.005$ to $0.010$.   Since it is difficult 
to reliably account for uncertainties in historic geological data,
let us extend our analysis through considering so-called `pink noise',
while assuming that $\sigma$  varies from $0.005$ to $0.009$.  Our
results are given in Table 2.  Our calculations show again that
the fractal dimension is not sensitive to the different values
of the noise and, moreover, that the WWS exhibits a small, but still
notable increasing trend with decreasing strength of noise $\sigma$.
To father explore this dependency, we consider pink noise with
$\sigma_{\cC}$, $\sigma_{\cM}$, and $\sigma_{\cO}$, corresponding
to the variables $\cC$, $\cM$, and $\cO$, respectivey.  Furthermore, 
we assume that $\sigma_{\cO}$ varies from $0.005$ to $0.009$, and
that  $\sigma_{\cC}$ = $\sigma_{\cM}$ = 0.005.  The obtained results
are given in Table 3, which shows that the trend of WWS is similar to
that attained in the last column of Table 2.

Having established the multifractal characteristics of the time series, 
the series has then been randomized and the WSS has been used to identify
the nature of its multifractality.  In general, multifractality can be either
due to a broad probability density function for the values of the time series,
or due to different long-range time correlations of the small and large 
fluctuations in the time series, or both.  The obtained results show
that our time series exhibit multifractality caused by long-range 
time correlations of the small and large fluctuations in the time series.
This is an important result, as it shows that the numerically generated 
time series of oxygen, carbon and methane still have some statistical 
correlations even if they are separated by very long time intervals.


\section{CONCLUSIONS}

In this study we expand on previous work aimed at modeling the rise of
oxygen in Earth's atmosphere encompassing both the Great Oxidation and
early stages of the gradual oxygen increase during the Proterozoic eon.
Experimentally, information on this process has been obtained through
extended biogeological studies.  This field of research is obviously of
significant importance for historical terrestrial studies, but it has
also potentially profound implications for the growing fields of
exobiolgy and planetary science.

In a previous paper, GLW06 presented a mathematical model for the
time-dependent behavior of carbon, methane and oxygen based on a set
of nonlinear differential equations.  This model incorporated numerous
geological, atmospheric, biological and biochemical processes, and was
able to reproduce the bifurcated nature of Earth's ancient atmosphere
consisting of a low-oxygen (less than $10^{-5}$ PAL) and a
high-oxygen state (more than $5 \times 10^{-3}$ PAL), including the
transition between the two.  This model was subsequently extended by
CRM09 who studied details of the switch between the two stages using
more general step functions for the underlying biogeological process.
The aim of this work was to explore whether or not the switch occurred
in jumps (yoyo-model) or exhibited a more gradual oxygen concentration
increase.  

In the present work we further augmented this model by investigating
the effects of the noise on the time variations of oxygen,
carbon and methane by using fractal and multifractal analysis techniques.
The overall goal was to offer a more realistic picture by accounting for
uncertainties in the geological data, which led us to add different
strengths of noise within the original equations describing Earth's
ancient atmosphere.
Our results confirm the relatively smooth oxygen increase previously
described by CRM09; however, we also consider the behavior of Earth's
atmospheric oxygen level in the time span following the Great Oxidation,
i.e., the early Proterozoic eon.  For that time span, the
obtained results show that the numerically generated time series
describing the time variations of oxygen, carbon and methane cannot
properly be described by a single fractal dimension because they exhibit
multifractal characteristics.  We also demonstrated that our time series
exhibit the multifractality caused by the long-range time correlations
of the small and large fluctuations in the time series.  We regard this
work as a further step toward elucidating the complex biogeochemical
processes of the early Earth, which gave rise to the previously identified
oxygen increase, among other dynamic developments.  Additional studies
may also require to consider more complex geodynamic and geobiological
models for the early Earth, which are currently developed by other research
groups.

\acknowledgments
We would like to thank the two reviewers for their valuable comments
that allowed us to significantly improve our original manuscript. 
This work has been supported in part by the SETI Institute (M. C.) as well
as by the Alexander von Humboldt Foundation (Z. E. M.) and by the University
of Texas at Arlington through its Faculty Development Program (Z. E. M.).



\begin{figure*}
\centering
\begin{tabular}{c}
\epsfig{file=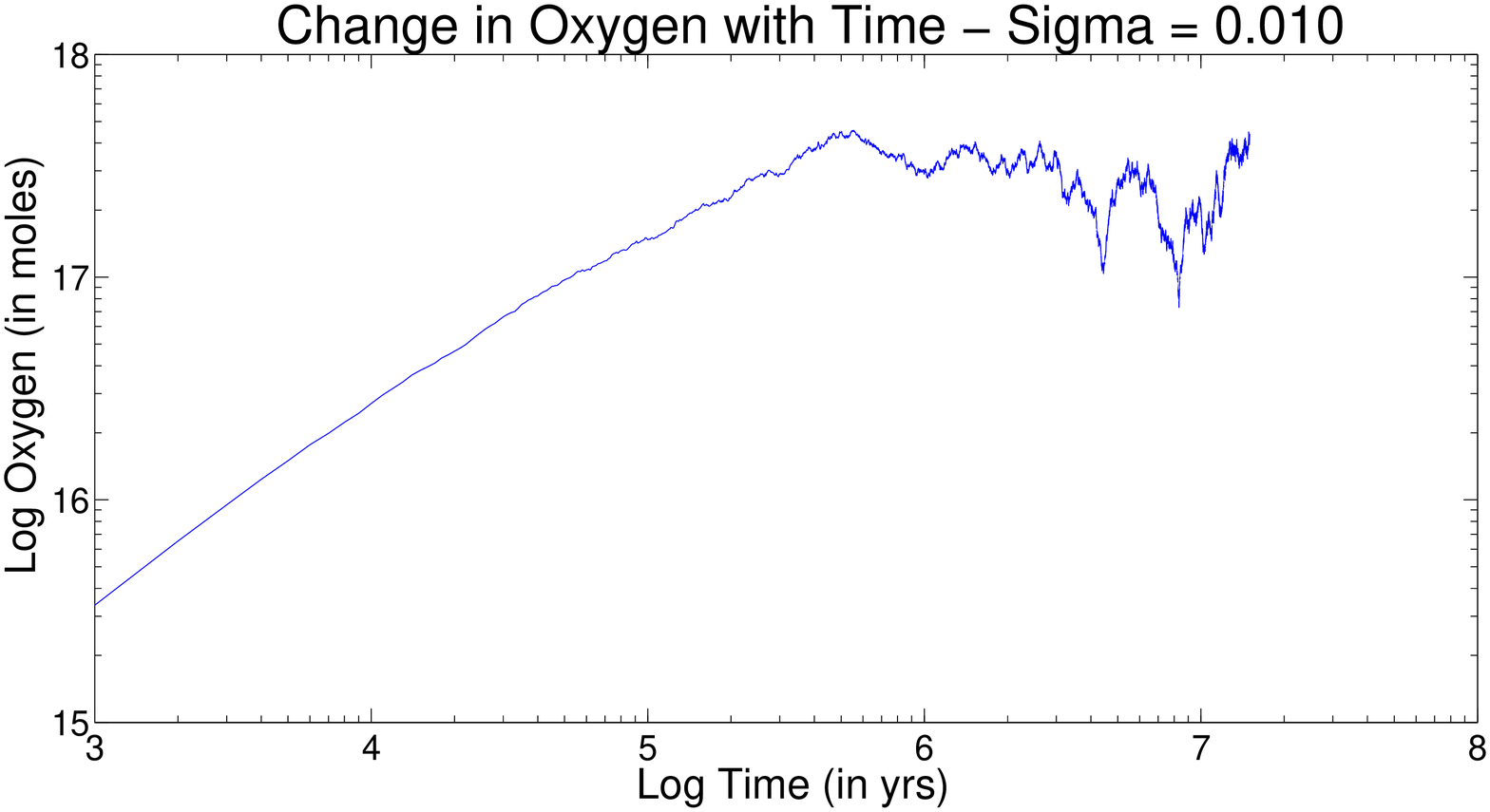,width=0.75\linewidth}
\end{tabular}
\caption{
Variations of oxygen with time in the early Earth's atmosphere
for a noise strength of $\sigma = 0.010$.  Note that
$3 \times 10^{17}$ mol of O$_2$ corresponds to 0.01 PAL.
}
\label{fig1}
\end{figure*}


\clearpage

\begin{figure*}
\centering
\begin{tabular}{c}
\epsfig{file=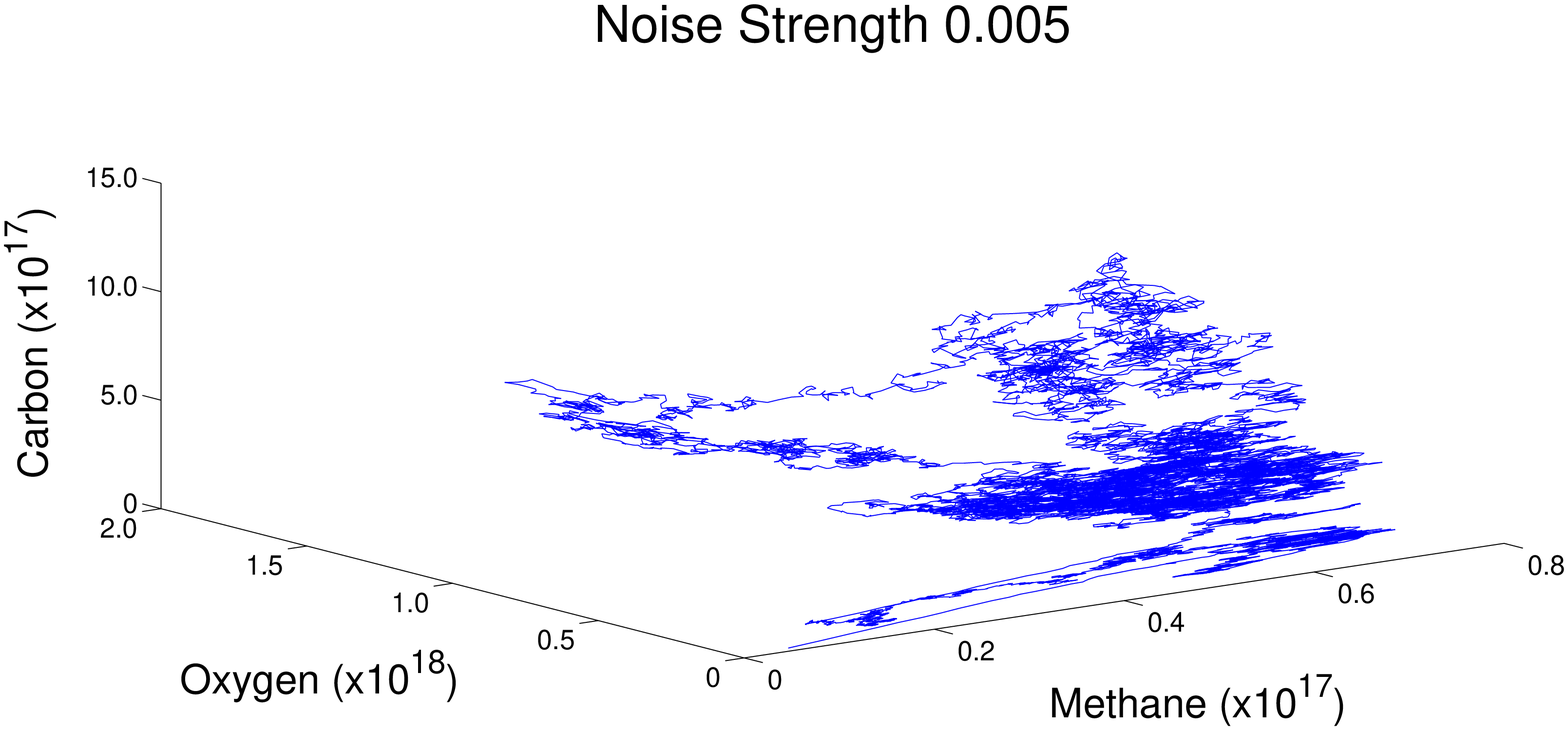,width=0.90\linewidth} \\
\epsfig{file=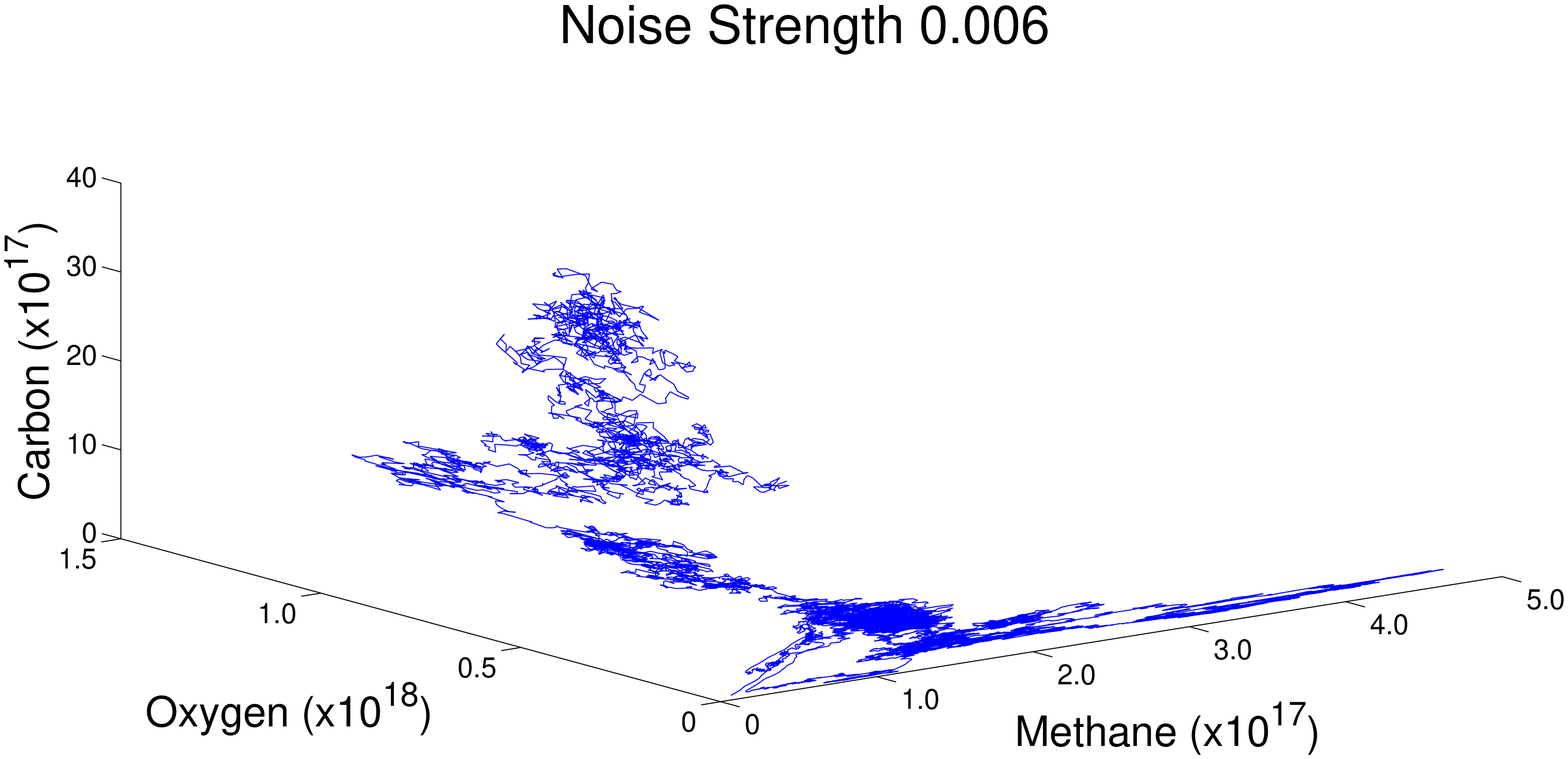,width=0.90\linewidth} \\
\epsfig{file=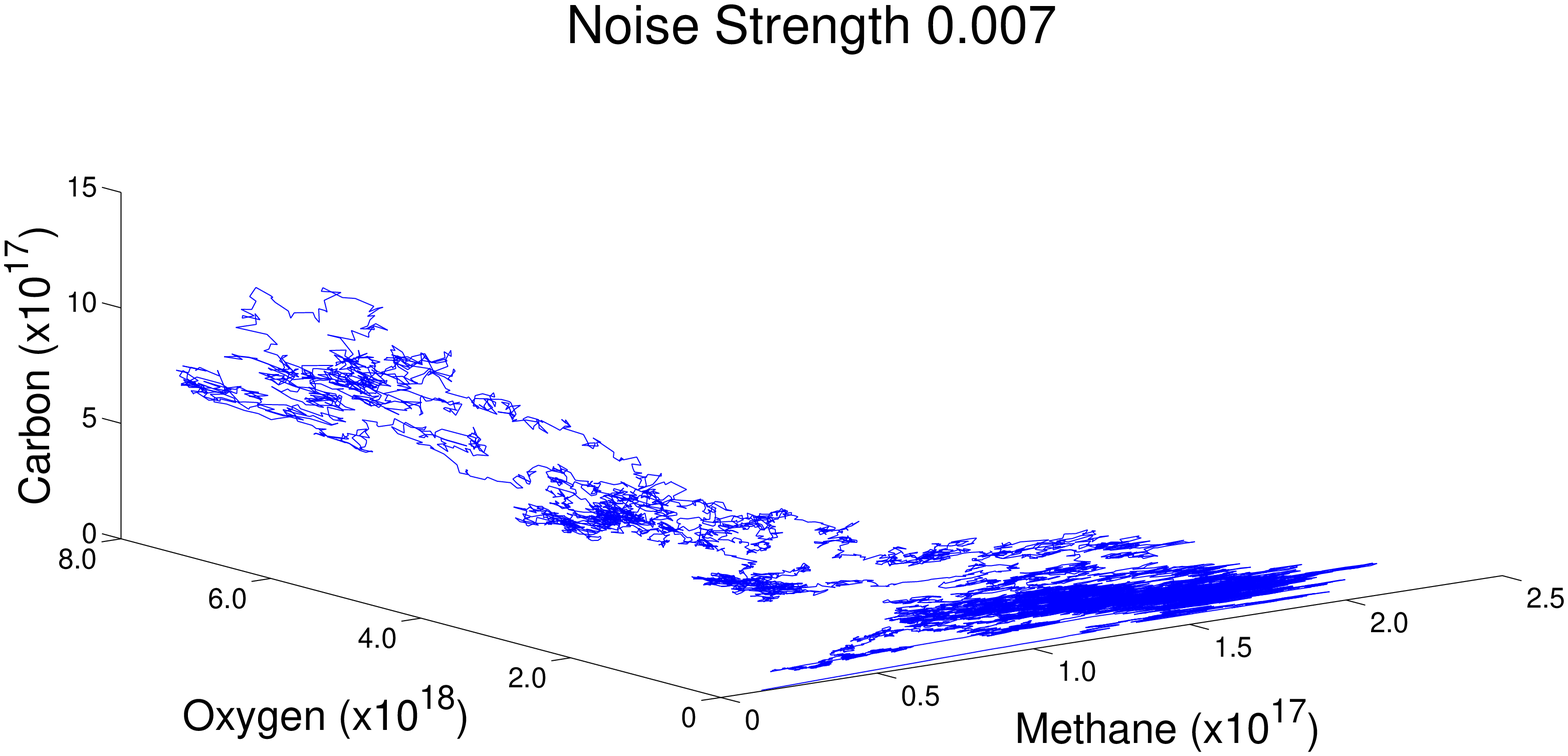,width=0.90\linewidth} 
\end{tabular}
\caption{
Phase portraits of our numerical results obtained for 
noise strengths of $\sigma = 0.005$, 0.006, and 0.007.
Note the differences in the various scales.
}
\label{fig2}
\end{figure*}


\clearpage

\begin{figure*}
\centering
\begin{tabular}{c}
\epsfig{file=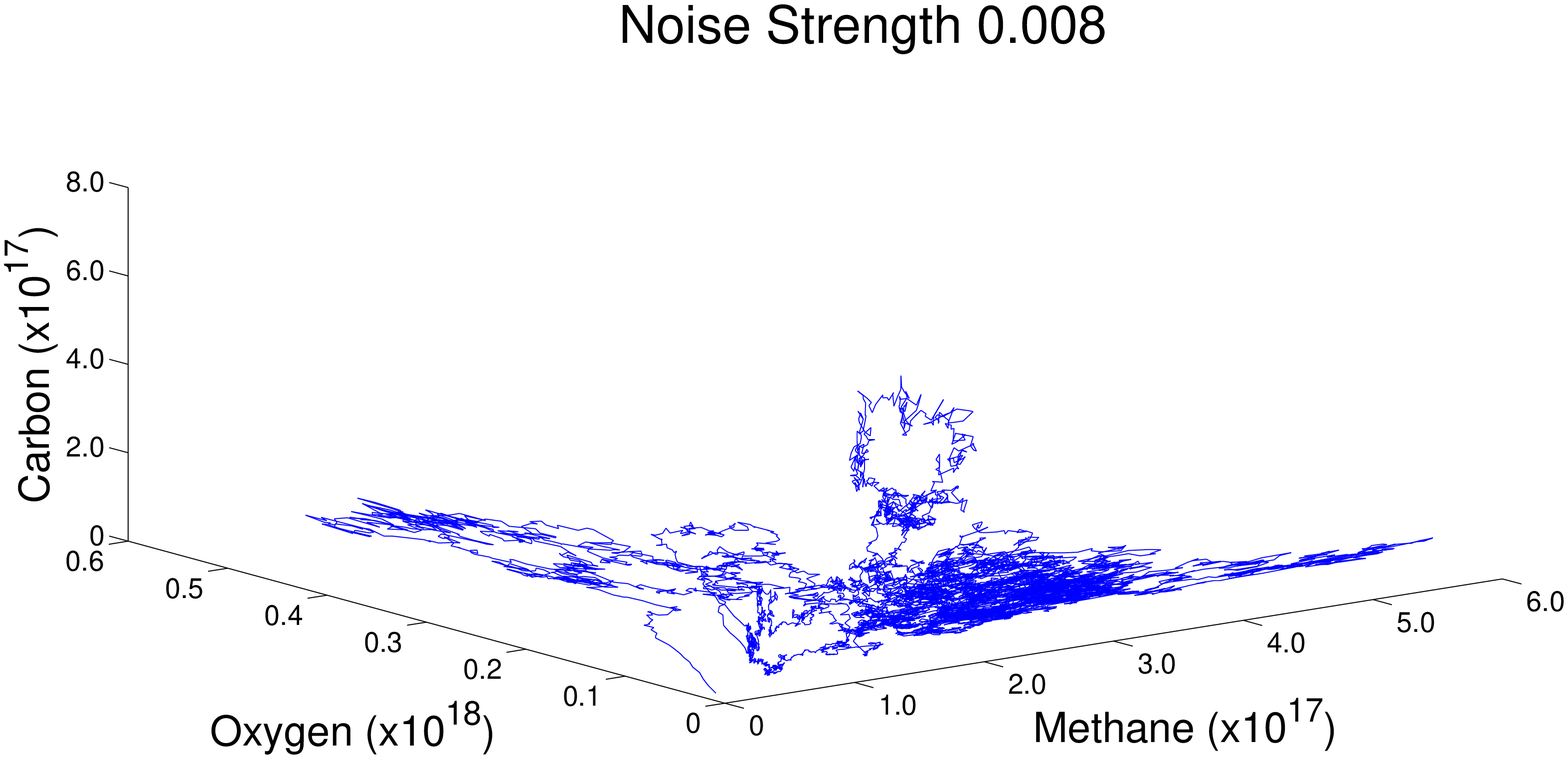,width=0.90\linewidth} \\
\epsfig{file=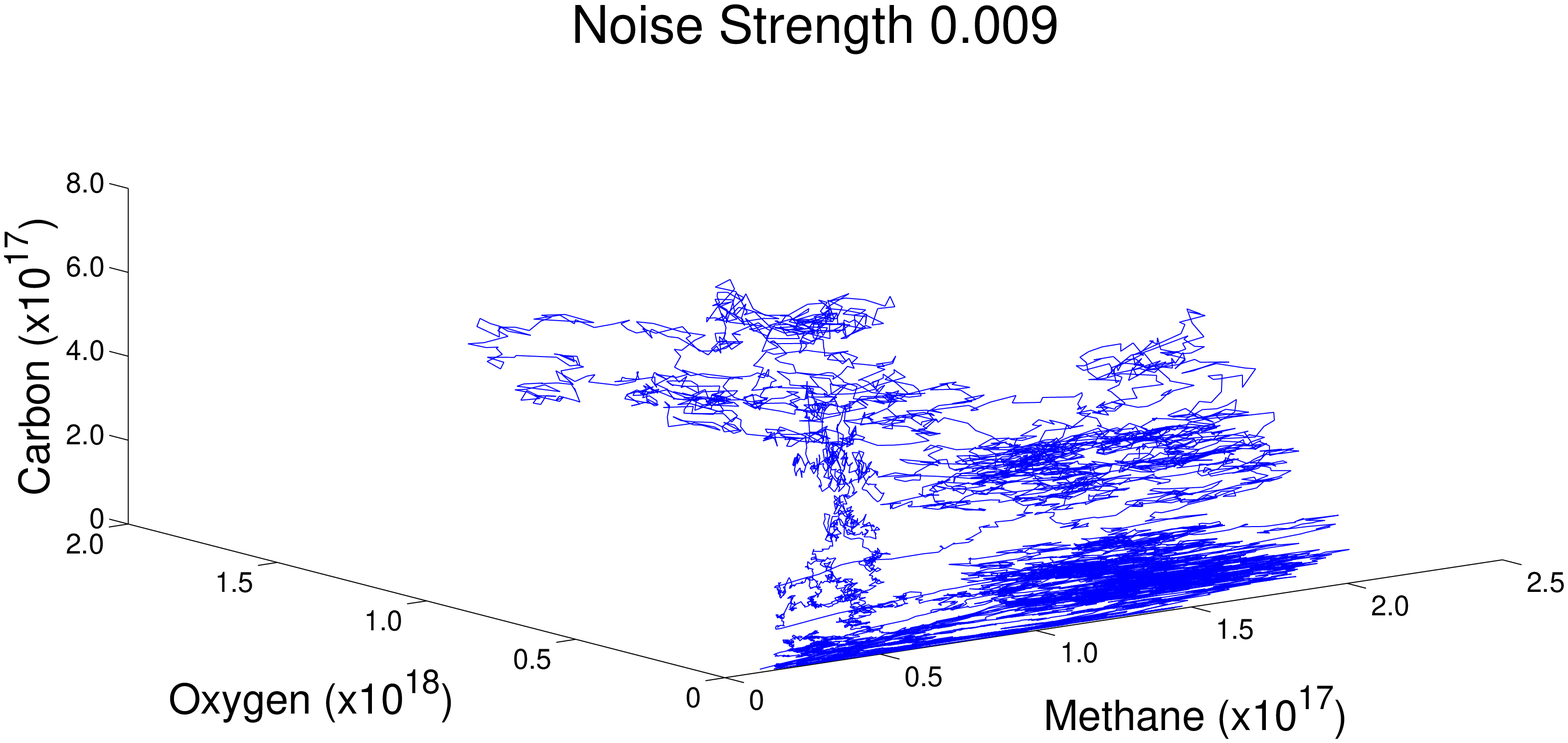,width=0.90\linewidth} \\
\epsfig{file=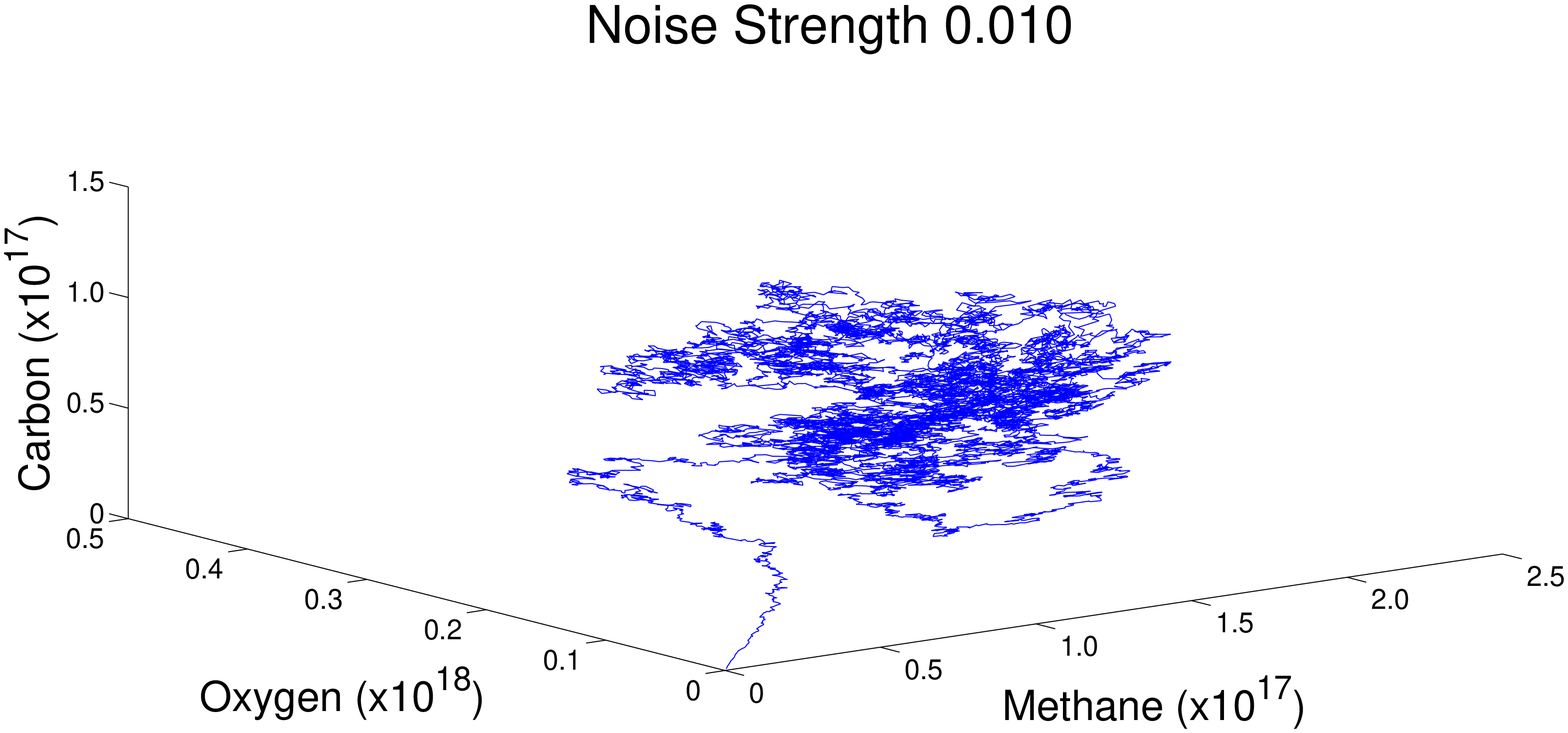,width=0.90\linewidth} 
\end{tabular}
\caption{
Phase portraits of our numerical results obtained for 
noise strengths of $\sigma = 0.008$, 0.009, and 0.010.
Note the differences in the various scales.
}
\label{fig3}
\end{figure*}


\clearpage


\begin{deluxetable}{cccc}
\tablecaption{Fractal and multifractal analysis of numerical results
with white Gaussian noise}
\tablewidth{0pt}
\tablehead{
    Noise strength & Generalized Hurst exponent & Fractal dimension & WSS  \\
    $\sigma$       & $h$ at $q = 2$             & $D_h$             & \omit      
}
\startdata
    0.005 & 0.5087 & 3.4913 & 1.2080 \\
    0.006 & 0.5104 & 3.4896 & 1.0760 \\
    0.007 & 0.5032 & 3.4968 & 1.1609 \\
    0.008 & 0.5457 & 3.4543 & 0.8647 \\
    0.009 & 0.5325 & 3.4675 & 1.0918 \\
    0.010 & 0.5206 & 3.4794 & 0.8110 \\   
\enddata
\label{tab:1}
\end{deluxetable}


\begin{deluxetable}{cccc}
\tablecaption{Fractal and multifractal analysis of numerical results
with pink noise}
\tablewidth{0pt}
\tablehead{
    Noise strength & Generalized Hurst exponent & Fractal dimension & WSS  \\
    $\sigma$       & $h$ at $q = 2$             & $D_h$             & \omit      
}
\startdata
    0.005 & 0.4496 & 3.5504 & 0.8196 \\
    0.006 & 0.4310 & 3.5690 & 0.8191 \\
    0.007 & 0.4486 & 3.5514 & 0.9876 \\
    0.008 & 0.4467 & 3.5533 & 0.8045 \\
    0.009 & 0.4361 & 3.5639 & 1.0165 \\
\enddata
\label{tab:2}
\end{deluxetable}


\begin{deluxetable}{cccc}
\tablecaption{Fractal and multifractal analysis of numerical results
with pink noise and $\sigma_{\cO}$ varying from $0.005$ to $0.009$,
and $\sigma_{\cC}$ = $\sigma_{\cM}$ = 0.005.}
\tablewidth{0pt}
\tablehead{
    Noise strength & Generalized Hurst exponent & Fractal dimension & WSS  \\
    $\sigma$       & $h$ at $q = 2$             & $D_h$             & \omit      
}
\startdata
    0.005 & 0.4496 & 3.5504 & 0.8196 \\
    0.006 & 0.4692 & 3.5308 & 0.6779 \\
    0.007 & 0.4786 & 3.5214 & 0.8045 \\
    0.008 & 0.4578 & 3.5422 & 0.7622 \\
    0.009 & 0.4582 & 3.5418 & 0.9527 \\
\enddata
\label{tab:3}
\end{deluxetable}

\end{document}